\documentclass[submission,copyright,creativecommons]{eptcs}
\providecommand{\event}{SYNT 2014} 
\usepackage{breakurl}             

\usepackage{amsmath,amsthm}
\usepackage{booktabs}
\usepackage{temporal}
\usepackage{enumerate}
\usepackage{graphicx}
\usepackage{pifont}
\usepackage{tikz}
\usetikzlibrary{calc,fadings}
\usepackage{xspace}

\newcommand{\Ass}{{\sf Ass}}
\newcommand{\Gua}{{\sf Gua}}

\newcommand{\impl}{~\rightarrow~}
\newcommand{\spec}{\phi}

\theoremstyle{definition}
\newtheorem{example}{Example}

\newcommand{\bc}{\emph{Be Correct}\xspace}
\newcommand{\dbl}{\emph{Don't Be Lazy}\xspace}
\newcommand{\ngu}{\emph{Never Give Up}\xspace}
\newcommand{\coop}{\emph{Cooperate}\xspace}
\newcommand{\exampleend}{\hfill $\star$}

\title{How to Handle Assumptions in Synthesis\thanks{This work was
supported in part by the Austrian Science Fund (FWF) through the
national research network RiSE (S11406-N23) and the project QUAINT
(I774-N23), as well as by the European Commission through project
STANCE (317753).}}
\author{Roderick Bloem$^1$ \qquad R\"udiger Ehlers$^{2,3}$ \qquad Swen Jacobs$^1$ \qquad Robert K\"onighofer$^1$\\
\institute{
\begin{tabular}{ccccc}
\multicolumn{1}{c}{$^1$Graz University of Technology} & \qquad \qquad & \multicolumn{1}{c}{$^2$University of Bremen} & \qquad \qquad & \multicolumn{1}{c}{$^3$DFKI GmbH}\\
 \multicolumn{1}{c}{Graz, Austria} & & \multicolumn{1}{c}{Bremen, Germany} & & \multicolumn{1}{c}{Bremen, Germany}
\end{tabular}}
}
\def\titlerunning{How to Handle Assumptions in Synthesis}
\def\authorrunning{R. Bloem, R. Ehlers, S. Jacobs \& R. K\"onighofer}
\begin{document}
\maketitle

\begin{abstract}
The increased interest in reactive synthesis over the last decade has led to
many improved solutions but also to many new questions. In this paper,
we discuss the question of how to deal with assumptions on environment
behavior. We present four goals that we think should be met and review
several different possibilities that have been proposed. We argue that
each of them falls short in at least one aspect.
\end{abstract}

\section{Introduction}
In reactive synthesis, we aim to automatically build a system that fulfills
guarantees $\Gua$ under the assumption that the environment fulfills
some properties $\Ass$. Most popular synthesis approaches take the
rudimentary view that the system and its environment are adversaries,
and that the synthesis problem is solved by generating a system that
realizes the formula $$\Ass \rightarrow \Gua.$$

We argue that this view is imperfect, describe
principles that we believe are important to obtain desirable systems,
review the work of others who have come to similar conclusions,
and describe drawbacks of the proposed approaches. The purpose of the
paper is to raise questions rather than to present answers, and to
highlight (our) lack of understanding of the problem, rather than our
understanding of a solution. Doing this, we hope to spark discussions and
further research on this topic.

To see that the setting described above is imperfect, consider a hypothetical example from
real life. Suppose that a coach promises the owners of his/her team to win a match
under the reasonable assumption that none of the coach's players gets injured
during the match. In order to fulfill this contract, the coach may either work
hard at winning the game, or may injure one of the players during the last few
minutes of the match. While the latter approach may not be unheard of\footnote{Fiorentina's coach (in 2012) and Nuova Cosenza's coach (in 2013)
mostly likely each had a different motivation for attacking their own players.},
it is generally frowned upon.
The same problem occurs in synthesis: a system may fulfill the
specification $\Ass \rightarrow \Gua$ by forcing the environment to
violate the assumptions, which is quite undesirable~\cite{KleinP10}.

We will assume that systems are implemented in a setting that consists
of multiple components. Some of these may be implemented by a
synthesis tool and are thus correct by construction. Some may be
implemented by a human and we should have good faith in correctness,
but not certainty. Some components involve physical interaction with the
environment, and we should be skeptical of the assumptions
that we have made about these components \cite{DBLP:conf/hybrid/TopcuOLM12}. The same applies for components
whose functionality is carried out by a human operator. 
Finally, note that even in a
perfectly implemented system, errors occur due to environmental
influences such as \emph{soft errors}~\cite{1479948}.

To be sure, the requirement that the system fulfills the guarantees in
all cases that the assumptions are fulfilled is very natural and
captures the notion of correctness. Yet, it is a very incomplete
notion of what is desirable in a system.
We present the following (non-exhaustive) list of functional goals that we believe a desirable system
should aim for:

\begin{description}
\item[Be Correct!] Fulfill the guarantees if the environment fulfills
  the assumptions.
\item[Don't be Lazy!] Fulfill the guarantees as well as possible for as many
  situations as possible, even when the assumptions are not
  fulfilled. 
\item[Never Give Up!] If you cannot satisfy the
  guarantees for every environment behavior, try to satisfy them when
  you can. 
\item[Cooperate!] If possible, allow or help the environment to
  fulfill the assumptions.
\end{description}
Note the difference between \emph{Don't be Lazy} and \emph{Never Give
  Up}: in the first case, we can enforce the guarantees. In the latter
we cannot enforce it, but we may be able to succeed if the environment
does not exhibit worst-case behavior.
Besides these four functional goals, we want the assumptions to be an
abstraction of the  environment's specifications so as not to make the
synthesis procedure unduly complex.

Traditional LTL synthesis only meets the goal to \emph{Be Correct}. In
Section~\ref{sec:correct}, we will show this in more detail, along with the
limitations of the approach. After that, we survey and illustrate some existing
approaches for the other goals. For \emph{Don't be Lazy} (Sect.~\ref{sec:lazy})
we focus on robust and error-resilient synthesis, and on synthesis with quantitative objectives
since these approaches attempt to satisfy guarantees as well as possible. For
\emph{Never Give Up} (Sect.~\ref{sec:giveup}) we will look at research that goes
beyond a purely adversarial view of games by suggesting reasonable strategies
for losing states. For \emph{Cooperate} (Sect.~\ref{sec:cooperate}) we will
leave the adversarial view even further:  We will consider non zero-sum
approaches which allow for explicit collaboration by constructing joint
strategies for the players. For each approach we will show how it addresses at
least one goal and how it is imperfect for another.  Finally, in
Section~\ref{sec:concl}, we will conclude our investigation with a table
summarizing the strengths and weaknesses of the discussed approaches.
\section{Be Correct!}
\label{sec:correct}
\subsection{Standard Synthesis}\label{sec:standard}
In standard  synthesis, the environment is treated as an adversary,
i.e., synthesized systems must be correct for \emph{any} possible
behavior of the environment. The behaviors of the environment under
which the guarantees $\Gua$ must hold are then modeled as antecedents
to the implication
%
\[ \Ass \impl \Gua. \] 
The corresponding payoff matrix\footnote{In all matrices, the absolute values
  of the payoffs are immaterial and only illustrate relative
  preferences.} is shown in
Figure~\ref{fig:matrix1}: the only one case that is considered undesirable is
when $\Ass$ is fulfilled, but $\Gua$ is not; there is no difference in
payoff or desirability between the three remaining cases.

\begin{figure}
\centering
\begin{tikzpicture}
\newcommand{\distance}{0.4}
\node[anchor=west] (gv) at (0,0) {$\Gua$ violated};
\node[anchor=north west] (gs) at ($(gv.south west)+(0,-\distance)$) {$\Gua$ satisfied};
\node[anchor=south west] (av) at ($(gv.north east)+(\distance,\distance)$) {$\Ass$ violated};
\node[anchor=west] (as) at ($(av.east)+(\distance,0)$) {$\Ass$ satisfied};

\path ($(gs.north west)+(-0.5*\distance,0.5*\distance)$) node (line3start) {} -| node (line3mid) {} ($(as.north west)+(-0.5*\distance,0.5*\distance)$) node (line3end) {};

\path ($(gv.north west)+(-0.5*\distance,0.5*\distance)$) node (line1start) {} -- ($(as.south east)+(0.5*\distance,-0.5*\distance)$) node (line1end) {};
\path ($(av.north west)+(-0.5*\distance,0.5*\distance)$)  node (line2start) {}  -- ($(gs.south east)+(0.5*\distance,-0.5*\distance)$) node (line2end) {};

\path (line1start) -| node (Q2UL) {} (line3end);
\path (line3start) -| node (Q2LR) {} (line1end);
\path[fill=black] (Q2UL) rectangle (Q2LR);

\draw[thick] ($(gv.north west)+(-0.5*\distance,0.5*\distance)$) -- ($(as.south east)+(0.5*\distance,-0.5*\distance)$);
\draw[thick] ($(av.north west)+(-0.5*\distance,0.5*\distance)$) -- ($(gs.south east)+(0.5*\distance,-0.5*\distance)$);

\path (gv) -| node {\textbf{1}} (av);
\path (gv) -| node {\color{white}\textbf{0}} (as);
\path (gs) -| node {\textbf{1}} (av);
\path (gs) -| node {\textbf{1}} (as);
\end{tikzpicture}
\caption{The standard desirability matrix for satisfying the specification.}
\label{fig:matrix1}
\end{figure}


First, the implication does not enforce \dbl: It does not
distinguish a trace that satisfies $\Gua$ from one that violates $\Ass$.
Thus, it does not restrict the behavior of the system in any way on
traces where the environment violates $\Ass$. (An example can be found
in Section~\ref{sec:lazy}.)

Second, the formalization does not imply the satisfaction of \ngu. Standard synthesis
only optimizes the worst-case behavior, i.e., if $\Ass \impl \Gua$
cannot be fulfilled for some behavior of the environment, then the
output of the synthesis algorithm will simply be ``unrealizable'',
instead of a system that fulfills $\Gua$ whenever possible. This goal
is all the more important in a situation in which the assumptions are
violated. In that case, the guarantees may not be realizable, but even
then they should be fulfilled whenever possible. (An example can be
found in Section~\ref{sec:giveup}.)

Third, the approach does not fulfill \coop: the system may
\emph{force} the environment to violate $\Ass$. 

\begin{example}\label{ex:forcefalse}
Consider the specification
\[ \left( \always \eventually r \land \always ( r \rightarrow \nextt (\neg r \weakuntil g)) \right) \impl \always ( r \rightarrow \nextt g). \]
This specification should result in a system that grants every request
in the next time step, for every environment that gives infinitely
many requests, but no request is repeated before there is a grant. In
this setting, requests are signaled by setting $r$ to $\true$, whereas
grants are signaled by setting $g$ to $\true$.  By simply giving no
grants at all (violating the guarantees), the system can force the
environment to violate the assumptions, thus fulfilling the
specification.\exampleend
\end{example}

The behavior shown in the example may be intended if the environment is considered to be purely 
adversarial, but in many applications of system design, this is not 
the case. Good examples for this fact are large systems that are constructed modularly, where 
the overall system is abstracted by environment assumptions for a particular 
component that we want to synthesize. Then our goal is not for the component 
to force the environment (i.e., the rest of the system) to violate the 
assumptions, but to work together with the environment to some extent, 
allowing both $\Ass$ and $\Gua$ to be satisfied whenever possible.

Thus, the standard approach fulfills \bc, but not \dbl, \ngu, and
\coop. This is summarized in Table~\ref{tab:overview} (on page \pageref{tab:overview}) together with
the strengths and weaknesses of the approaches discussed in the next sections. 

\section{Don't Be Lazy!}
\label{sec:lazy}

Traditionally, correctness is considered to be a Boolean property: a system either
realizes a specification or not.  For specifications of the form $\Ass
\rightarrow \Gua$, this attitude results in the desirability matrix shown in
Figure~\ref{fig:matrix1}.  This section focuses on improving the system
behavior if assumptions are violated, i.e., on the left column of the matrix.

\begin{example}
\label{ex:flight0}
As motivation, consider a flight control system which must work correctly under
the assumption that the number of simultaneously arriving planes is less than
$100$.  For more planes, the specification may be unrealizable, e.g., because it
may be impossible to guarantee all timing constraints.  Suppose further that the
system has been synthesized, is in operation, and the $101^\text{st}$ plane
arrives.  A work-to-rule synthesis algorithm could have considered this
situation as \emph{won}, and may have randomly chosen to stop serving \emph{any}
plane in this situation.  A more desirable system would serve planes as well as
possible, even though the assumption is violated: For instance, ignoring the
$101^\text{st}$ plane or responding a bit slower are certainly better options.
Even more, for configurations of the $101$ planes that can be handled with
the available resources, it would be preferable if no reduction in the quality
of service occurs at all. \exampleend
\end{example}

With the matrix in Figure~\ref{fig:matrix1}, once the assumptions are violated, there
is no additional benefit for the system to satisfy the guarantees any more.  
The
implication is satisfied for any future system behavior, so it can then behave arbitrarily. 
The synthesis
algorithm can exploit this freedom even in situations in which it would still be
possible to satisfy the guarantees. This is clearly undesirable. Intuitively,
the synthesized system should always aim for satisfying the guarantees, even if
assumptions are violated, instead of getting lazy and doing only the least to
satisfy the implication.

With the payoff matrix in Figure~\ref{fig:matrix_robust}, this changes. By
giving traces of the system in which the system satisfies the guarantees a
higher payoff regardless of whether the assumptions are satisfied, there is
always an incentive for the system to satisfy the guarantees. An approach to
deal with multiple ranked specifications is presented in
\cite{DBLP:conf/cav/AlurKW08}. 

In practice, assumptions and guarantees can be violated only slightly, or very
badly. With this non-Boolean understanding of property violations, the
desirability matrix of Figure~\ref{fig:matrix_robust} gets blurred, with gradual
transitions between the quadrants, as represented in
Figure~\ref{fig:matrixBlurred}.  It makes sense to consider the degree in which
guarantees are satisfied also in synthesis: even if it is not possible to
satisfy all guarantees due to assumption violations, an ideal system would still
try to satisfy guarantees ``as well as possible''.

In the remainder of this section, we briefly review previous approaches to
synthesize systems that are eager to satisfy their guarantees.
We start by reviewing the \emph{strict implication semantics} employed in the \emph{Generalized Reactivity(1) Synthesis} approach in Section~\ref{sec:gr1},
which yields a form of such eagerness as a by-product.
In Section~\ref{sec:error_res}, we then discuss approaches that extend the set of environment behaviors under
which the system can satisfy its guarantees, and in this way make the system less
lazy without sacrificing the satisfaction of the guarantees. Synthesis approaches that allow slight deviations from the guarantees in case of assumption
violations are discussed in Section~\ref{subsec:robust}.
Finally, we discuss quantitative synthesis in Section~\ref{sec:quant}, which offers a flexible framework
to encode quality criteria of synthesized systems, including some notions of
eagerness of the system to satisfy its guarantees.

\begin{figure}
\centering
\begin{tikzpicture}
\newcommand{\distance}{0.4}
\node[anchor=west] (gv) at (0,0) {$\Gua$ violated};
\node[anchor=north west] (gs) at ($(gv.south west)+(0,-\distance)$) {$\Gua$ satisfied};
\node[anchor=south west] (av) at ($(gv.north east)+(\distance,\distance)$) {$\Ass$ violated};
\node[anchor=west] (as) at ($(av.east)+(\distance,0)$) {$\Ass$ satisfied};

\path ($(gs.north west)+(-0.5*\distance,0.5*\distance)$) node (line3start) {} -| node (line3mid) {} ($(as.north west)+(-0.5*\distance,0.5*\distance)$) node (line3end) {};

\path ($(gv.north west)+(-0.5*\distance,0.5*\distance)$) node (line1start) {} -- ($(as.south east)+(0.5*\distance,-0.5*\distance)$) node (line1end) {};
\path ($(av.north west)+(-0.5*\distance,0.5*\distance)$)  node (line2start) {}  -- ($(gs.south east)+(0.5*\distance,-0.5*\distance)$) node (line2end) {};

\path (line1start) -| node (Q2UL) {} (line3end);
\path (line3start) -| node (Q2LR) {} (line1end);
\path[fill=black] (Q2UL) rectangle (Q2LR);

\draw[thick] ($(gv.north west)+(-0.5*\distance,0.5*\distance)$) -- ($(as.south east)+(0.5*\distance,-0.5*\distance)$);
\draw[thick] ($(av.north west)+(-0.5*\distance,0.5*\distance)$) -- ($(gs.south east)+(0.5*\distance,-0.5*\distance)$);

\path (gv) -| node {\textbf{1}} (av);
\path (gv) -| node {\color{white}\textbf{0}} (as);
\path (gs) -| node {\textbf{2}} (av);
\path (gs) -| node {\textbf{2}} (as);
\end{tikzpicture}
\caption{The desirability matrix used in error-resilient synthesis.}
\label{fig:matrix_robust}
\end{figure}

\begin{figure}
\pgfdeclarehorizontalshading{myshadingA}{100bp}
{
color(0bp)=(pgftransparent!100);
color(25bp)=(pgftransparent!100);
color(50bp)=(pgftransparent!100);
color(85bp)=(pgftransparent!0);
color(100bp)=(pgftransparent!0)
}
\pgfdeclarefading{fadingmyS}{\pgfuseshading{myshadingA}}
\centering
\begin{tikzpicture}
\newcommand{\distance}{0.4}
\node[anchor=west] (gv) at (0,0) {$\Gua$ violated};
\node[anchor=north west] (gs) at ($(gv.south west)+(0,-\distance)$) {$\Gua$ satisfied};
\node[anchor=south west] (av) at ($(gv.north east)+(\distance,\distance)$) {$\Ass$ violated};
\node[anchor=west] (as) at ($(av.east)+(\distance,0)$) {$\Ass$ satisfied};

\path ($(gs.north west)+(-0.5*\distance,0.5*\distance)$) node (line3start) {} -| node (line3mid) {} ($(as.north west)+(-0.5*\distance,0.5*\distance)$) node (line3end) {};

\path ($(gv.north west)+(-0.5*\distance,0.5*\distance)$) node (line1start) {} -- ($(as.south east)+(0.5*\distance,-0.5*\distance)$) node (line1end) {};
\path ($(av.north west)+(-0.5*\distance,0.5*\distance)$)  node (line2start) {}  -- ($(gs.south east)+(0.5*\distance,-0.5*\distance)$) node (line2end) {};

\path (line1start) -| node (Q1OL) {} (line2start);
\path (line2start) |- node (Q1UL) {} (line3mid);
\path (line2end) -| node (Q3UR) {} (line3mid);
\path (line3start) -| node (Q2LR) {} (line1end);
\path (line1end) |- node (Q4LR) {} (line2end);

\path[line width=0pt,fill=black,path fading=west] (Q1OL) rectangle (line3mid);

\path[line width=0pt,fill=black,path fading=south] (line3mid) rectangle (Q4LR);

\path[line width=0pt,fill=black] (line3mid) rectangle (line1end);

\path[line width=0pt,fill=black,path fading=fadingmyS,fading angle=45] (line2end) rectangle (line3mid);



\draw[thick] ($(gv.north west)+(-0.5*\distance,0.5*\distance)$) -- ($(as.south east)+(0.5*\distance,-0.5*\distance)$);
\draw[thick] ($(av.north west)+(-0.5*\distance,0.5*\distance)$) -- ($(gs.south east)+(0.5*\distance,-0.5*\distance)$);

\path (gv) -| node {\textbf{1}} (av);
\path (gv) -| node {\color{white}\textbf{0}} (as);
\path (gs) -| node {\textbf{2}} (av);
\path (gs) -| node {\textbf{2}} (as);
\end{tikzpicture}
\caption{The (blurred) desirability matrix used in robust synthesis.}
\label{fig:matrixBlurred}
\end{figure}

\subsection{Assumptions in Generalized Reactivity Games}
\label{sec:gr1}
Specifications in the generalized reactivity fragment of rank 1 (GR(1)) have
been proposed as an alternative to full LTL, as their synthesis problems are
efficiently decidable and are still sufficiently expressive for many important
properties \cite{Bloem12}. What is particularly interesting in our present
comparison is that in GR(1) synthesis games that solve the synthesis problem for
this fragment, the implication $\Ass \impl \Gua$ is interpreted slightly
different than in the standard semantics (see Bloem et al.\ \cite{Bloem12} and
Klein and Pnueli~\cite{KleinP10}). In particular, safety guarantees and
assumptions are treated differently: even if the environment does not satisfy
$\Ass$ completely, the system must satisfy its safety guarantees at least as
long as the environment satisfies the safety assumptions. This rules out some
non-intuitive behavior by the system, where it violates $\Gua$ because it knows that
it can force the environment to violate $\Ass$ at some point in the future.  In
particular, the unintended behavior in Example~\ref{ex:forcefalse} is ruled out.


While this changes the \emph{rules} of the synthesis game such that the system
player loses the game if such a safety guarantee is violated before the environment violates some safety assumption (instead of
the system winning whenever the environment violates $\Ass$ anywhere in the
infinite trace), it does not change the purely adversarial view on the game.
\begin{example}
If the safety guarantee $\always ( r \rightarrow \nextt g)$ in
Example~\ref{ex:forcefalse} is changed into a liveness guarantee $\always ( r
\rightarrow \eventually g)$, i.e., the specification is modified to
\[ \left( \always \eventually r \land \always ( r \rightarrow \nextt (\neg r
\weakuntil g)) \right) \impl \always ( r \rightarrow \eventually g), \]
the system can still enforce an assumption violation by violating
the guarantee, even in the modified semantics of the implication, by not
giving any grants. The reason is that the system does not violate the guarantee
\emph{before} the assumption is violated. \exampleend
\end{example}
Furthermore, this extension does not change the purely worst-case analysis that
will simply return ``unrealizable'' if the specification cannot be fulfilled in
all cases, and otherwise return a solution that does not distinguish between
cases where $\neg \Ass \land \neg \Gua$ holds versus cases where $\Gua$ is
actually satisfied.
(Recall that strengths and weaknesses are summarized in
Table~\ref{tab:overview}.)

Related mechanisms are presented in~\cite{DIppolitoBPU13}.  This work presents
an approach to synthesize event-based behavior models from GR(1)
specifications. It uses the following definitions in order to avoid systems that
satisfy the specification by violating assumptions.  A \emph{best effort system}
system satisfies the following condition: if the system forces $\Ass$ not to
hold after a finite trace $\sigma$, then no other system that achieves $\Gua$
could have allowed $\Ass$ after $\sigma$.  An even stronger definition is that
of an \emph{assumption preserving} system: the system should never
prevent the environment from fulfilling its assumptions.  Every assumption
preserving system is also a best effort system.  Finally, the authors propose
\emph{assumption compatibility} as a methodological guideline. It is a
sufficient condition under which any synthesized system is assumption
preserving: The environment must be capable of achieving $\Ass$ regardless of
system behavior.  This can be checked by deciding
realizability with swapped roles.  However, this condition is rather strong.

\subsection{Synthesizing Error-Resilient Systems}
\label{sec:error_res}

The most desirable form of the system to react to environment assumption violations is
to continue to satisfy its guarantees. As in a system engineering process, assumptions
are typically only added on an as-needed basis, this will only be possible in rare
circumstances, and the synthesized system can then simply be made robust against
assumption violations by removing them before performing synthesis.

Yet, this does not mean that every single assumption violation requires the system
to violate its guarantees. A couple of approaches aim at exploiting this fact.

Topcu et al.~\cite{DBLP:conf/hybrid/TopcuOLM12} describe an approach to weaken
the safety part of the assumptions as much as possible in context of GR(1) synthesis.
The weakening is performed in a very fine-grained way, much finer than how a
human specifier would do so, and as fine-grained as possible in GR(1) synthesis without
the introduction of additional output signals to encode more complex properties.
The resulting synthesized controller is then completely error-resilient against 
environment behavior that is forbidden by the original assumptions,
but allowed by the refined assumptions.

Ehlers and Topcu~\cite{DBLP:conf/hybrid/EhlersT14} approach the problem from a
different angle. They describe how to synthesize a $k$-resilient implementation.
The notion of $k$-resilience has been defined earlier by Huang et al.~\cite{DBLP:journals/corr/abs-1210-2449}. 
Adapted to the case of GR(1) specifications, it requires
the system to satisfy the guarantees if not more than $k$ safety assumption violations
occur in between assumption-violation-free periods of the system execution,
provided that these
periods are long enough to allow the system to recover. The approach also allows
a more fine-grained analysis of for which assumptions some of their violations
can be tolerated and for which no violation can be tolerated -- whenever there is a trade-off
between the choices of assumptions for which violations should be tolerated, all
Pareto-optimal such choices are presented to the specifier.

Orthogonal to $k$-resilient synthesis is the idea to extend
a synthesized implementation  by \emph{recovery transitions} \cite{WEKG2014RSS}. Such transitions can be added for cases in which the assumptions are violated, but for which the system can react in a way that does not jeopardize the system's ability to completely satisfy its guarantees along its run if the environment starts to satisfy its assumptions again. In contrast to $k$-resilient synthesis, recovery behavior is added on a best-effort basis and the synthesized system does not strategically choose its nominal-case behavior such that as many safety assumption violations as possible are tolerated.

All three approaches only make the system robust to a certain extend
as they extend the set of environment behaviors under which a system can be synthesized. They
do not help to satisfy the guarantees as well as possible for environment behaviors
that do not fall into this set.

\subsection{Synthesis of Robust Systems}
\label{subsec:robust}

The basic idea of robust synthesis is to satisfy guarantees as
well as possible, even if assumptions are violated. Slight violations of the
guarantees are allowed when the assumptions are violated, and we can further
distinguish between different severity levels of assumption- and guarantee
violations.

Robust synthesis is motivated by the observation that synthesized systems
sometimes simply stop responding in any useful way after an assumption has been
violated. Consider the following example.

\begin{example}
\label{ex:arb0}
A system must grant two requests, but not simultaneously:
$\Gua = \always((r_0 \rightarrow g_0) \wedge (r_1 \rightarrow g_1) \wedge
(\neg g_0 \vee \neg g_1))$.
The environment must not raise both requests simultaneously: $\Ass =
\always(\neg r_0 \vee \neg r_1)$.  The plain implication $\Ass \rightarrow
\Gua$ allows the system to ignore any future request if the environment ever
happens to raise both requests. Optimizations for other properties
like circuit size of the synthesized solution may exploit this freedom.  However, a system that ignores one
of the simultaneous requests and then continues normally instead of getting lazy
would be more preferable. \exampleend
\end{example}

Of course, in case of violated assumptions, it may not always be possible to
satisfy all guarantees, as Example~\ref{ex:arb0} shows.  Otherwise, some assumptions would be superfluous.  Also,
it makes sense to take the severity of the assumption violation into account.
Intuitively, a small assumption violation should also lead to only small
guarantee violations.  Therefore, the crux in robust synthesis is to define
measures of how well guarantees are satisfied and how severe assumptions are
violated.  Then, an optimal ratio with respect to these metrics can be enforced.
Existing approaches~\cite{Bloem13} typically optimize the worst case of this ratio.
For safety properties, a natural conformance measure for both assumptions and
guarantees is to count the number of time steps in which properties are
violated. 
For liveness properties, this does not work because a liveness
property violation cannot be detected at any point in time: If some event is
supposed to happen eventually, and has not happened yet, we may just not have
waited long enough.
If $\Ass$ and $\Gua$ are composed of
several properties, one can also count the number of violated properties to
define the severity of a violation~\cite{Bloem13}. 

Despite the fact that liveness assumption violations cannot be observed at runtime, 
robust synthesis approaches for specifications with liveness assumptions and
guarantees exist that let the system tolerate (safety) assumption violations. 
Intuitively, the idea is to ask the system to tolerate safety assumption violations
if in only finitely many steps of the system's execution, such violations
occur. The system is then only allowed to violate safety guarantees finitely often.
Liveness assumptions are assumed to hold at all times. Since the system cannot know when an
assumption violation has been the last one, it has to behave in a robust way
\cite{DBLP:conf/nfm/Ehlers11}. As a variant to the approach, the system can additionally
be required to satisfy the liveness guarantees even if safety assumptions are
violated infinitely often \cite{DBLP:journals/corr/abs-1207-1268}.

In summary, robustness is definitely a useful extension to correctness.  One
shortcoming of existing solutions is that they only optimize the robustness
measure for the worst case, i.e., assume a perfectly antagonistic environment.
As a consequence, the resulting system may still be unnecessarily lazy for more
cooperative environment behaviors.  The fact that the system cannot satisfy the
guarantees any better in the worst case should not be an excuse for not trying.
In this sense, not assuming a fully adversarial environment in the robustness
optimization may yield even better results.  This aspect will be elaborated in
Section~\ref{sec:giveup}.

\subsection{Quantitative Synthesis}
\label{sec:quant}

Among all systems that realize a given specification, some may be more desirable
than others.  The idea of synthesis with quantitative objectives
is to construct a system that not only satisfies the (qualitative)
specification, but also maximizes a (quantitative) desirability metric.  In this
sense, some approaches to robust synthesis, as discussed in the previous section, can be seen as
special cases of quantitative synthesis.  But quantitative synthesis can also be
a handy tool to optimize solutions with respect to other desirability
metrics.

\begin{example}
Continuing Example~\ref{ex:arb0}, we may prefer systems that give as few
unnecessary grants as possible.  This can be achieved by assigning costs to
unnecessary grants (i.e., situations with $g_i \wedge \neg r_i$), and let the
synthesis algorithm minimize these costs. \exampleend
\end{example}

Of course, one could also specify each and every situation where no grant should
be given.  While this is quite possible for this small example, it can be
tedious, error-prone, and destroy the abstract quality of the specification for
more complicated cases:  Ideally, a specification only expresses \emph{what} the
system should do, but not \emph{how}.  If the exact behavior needs to be
specified for each and every situation, it is better to implement the system
right away.

The work of~\cite{Bloem09} presents a machinery based on games with a
lexicographic mean-payoff objective and a simultaneously considered parity
objective to solve such problems.  The parity objective encodes the qualitative
specification, while the mean-payoff objective encodes the quantitative
desirability metric.  The approach assumes fully adversarial environments and
optimizes for this worst case.

Defining a desirability metric for a system is never an easy task. Cern{\'y}
and Henzinger~\cite{Cerny11} propose to define it in two steps.  The first step
is to assign costs (or payoffs) to single traces.  This can be done by combining
the costs of single events in the trace, e.g., by taking the sum, average,
maximum, etc. Second, the costs for individual traces are combined into
total costs. Again, there are various options like taking the worst case, the
average case, or a weighted average assuming some probability distribution.
Although this approach is quite generic, it is questionable if the desirability
of a system can be expressed by one single number in the venture of satisfying
guarantees as good as possible in as many situations as possible.  Dominance
relations inducing a partial order between systems, as used in the next section,
may be a more natural notion as they provide a natural quantification over
environment behaviors.

If cost notions for both the environment and system actions can be given, there
is a canonical way to define which system traces are desirable: the ones that
are the cheapest. Tabuada et al.~\cite{DBLP:conf/emsoft/TabuadaBCSM12} adapt
notions from control theory to define a preferability relation on system
behaviors. In addition to minimizing the ratio between environment behavior cost
and system behavior cost, they also require that the effect of sporadic
disturbances vanishes over time.

Finally, there are approaches that combine quantitative approaches with a
\emph{probabilistic} model of the environment, to find the best solution under a
given probability distribution for actions of the environment~\cite{EssenJ12}. A
combination of probabilistic and worst-case reasoning is considered by Bruy\`ere
et al.~\cite{BruyereFRR14}.

In summary, quantitative synthesis does not directly address the problem of
dealing with assumptions in synthesis, but can rather be seen as a tool
for obtaining better solutions with respect to different metrics.  The fact that
the environment is considered as perfectly adversarial in most methods may
not be ideal in all settings.


\section{Never Give Up!}
\label{sec:giveup}



\noindent
Traditional games-based synthesis is only concerned with the worst case.  As
already raised in the previous sections, this mind-set is not always justified.

\begin{example}
The flight control system from Example~\ref{ex:flight0} may actually be able to
handle way more than 100 planes in time if they do not all signal an emergency
at the same time. This worst case is possible, but very unlikely to happen in
practice. \exampleend
\end{example}

If a guarantee cannot be enforced in the worst case, traditional synthesis
methods will consider this guarantee as ``impossible'' to achieve.  Thus, the
constructed system would behave arbitrary if it ever gets into such a
``hopeless'' situation, i.e., it would not even try to reach the goal.  However,
when the system is in operation, its concrete environment may not be perfectly
adversarial, i.e., the worst case may not occur.  Hence, it makes sense for the
system to behave faithfully even in (worst-case-)lost situations instead of
resigning. In other words, the synthesized system should retain or even maximize
the chances of reaching the goal (e.g., satisfying all guarantees even if
assumptions are violated), even if this is not possible in the worst case.

Note the difference to robust and quantitative synthesis, as discussed in the
previous section: Robust and quantitative synthesis aim at satisfying guarantees
as well as possible for the worst case environment behavior.  In contrast, this
section is concerned with satisfying the specification (preferably without cut-backs) for
many environment behaviors that do not represent the worst case as they violate $\Ass$.
In the following, we will discuss existing synthesis approaches that tackle this
problem by dealing with ``hopeless'' situations in a constructive way.


\subsection{Environment Assumptions}
\label{sec:environment-assumptions}
When we consider the basic idea of ``restricting'' the environment behavior 
by adding assumptions to an LTL specification of the form $\Ass \impl 
\Gua$, then synthesis from such a specification results in a system that is 
guaranteed to satisfy $\Gua$ for all behaviors of the environment that 
satisfy $\Ass$. On the other hand, the system does not give any guarantees 
for traces on which $\Ass$ is violated.

Chatterjee, Henzinger and Jobstmann~\cite{Chatter08} show how, for a given
unrealizable system specification $\Gua$, one can compute an environment
assumption $\Ass$, such that $\Ass \impl \Gua$ is realizable (for
$\omega$-regular specifications). The computed assumptions consist of a safety
and a liveness part, and should be as weak as possible. While
\emph{minimal} (but not unique) safety assumptions\footnote{\emph{Minimal} here
means that a minimal number of environment edges are removed from the game
graph.} can be computed efficiently, the problem is NP-hard for minimal liveness
assumptions\footnote{Here, \emph{minimal} means to put fairness conditions on
a minimum number of environment edges in the game graph.}. If it is sufficient
to compute a \emph{locally minimal} set of liveness assumptions, i.e., a set of
liveness assumptions from which no element may be removed without changing the
resulting specification to be realizable, NP-hardness can be avoided.

\subsection{Best-Effort Strategies for Losing States}
\label{sec:best_effort}

Faella~\cite{Faella08,FaellaWorkshopPaper} investigates best-effort strategies for states from which
the winning condition cannot be enforced.  Intuitively, a good strategy should
behave rationally in the sense that it does not ``give up''.  Hence, this work
assumes the desirability matrix of Figure~\ref{fig:matrix1}, and is concerned
with staying away from the top-right corner, even if this is not possible in the
worst case.

\begin{example}
As an example, consider the specification $\always\eventually(o \wedge
\nextt i)$,
where $i$ is an input and $o$ is an output.  There is no way the system can
enforce satisfying the property.  However, setting $o$ to $\true$ as often as
possible is more promising than setting $o$ always to $\false$. \exampleend
\end{example}

Faella~\cite{Faella08} discusses and compares several goal independent criteria
for such rational strategies.  The work concludes that \emph{admissible}
strategies, defined via a dominance relation, may be a good choice. Intuitively,
strategy $\sigma$ \emph{dominates} strategy $\sigma'$ if $\sigma$ is always at
least as good as $\sigma'$, and better for at least one case. More specifically,
$\sigma$ dominates $\sigma'$ if (1) for all environment strategies and starting
states, if $\sigma'$ satisfies the specification then $\sigma$ does so too, and
(2) there exists some environment strategy and starting state from which
$\sigma$ satisfies the specification but $\sigma'$ does not.  This induces a
partial order between strategies.  An \emph{admissible} strategy is one that is
not dominated by any other strategy.

For positional\footnote{A goal is positional if the strategy does not require
memory on top of knowing the current position in \emph{synthesis games} that are built from the given specification.} and prefix-independent\footnote{A goal is prefix-independent if
adding or removing a finite prefix to/from the execution does not render a
satisfied property violated.} goals, Faella~\cite{Faella08} presents an
efficient way to compute admissible strategies: the
conventional winning strategy $\sigma$ is computed and played from all winning
states.  For the remaining states, a cooperatively winning strategy $\sigma'$ is
computed, assuming that $\sigma$ is played in the winning states.  This is
a very relevant result because, e.g., parity goals are positional and
prefix-independent, and LTL specifications can be transformed into parity games.
 For goals that do not fall into this category, the computation of admissible
strategies is left for future work.  Unfortunately, this work has not been
actively followed up on.

Damm and Finkbeiner also consider admissible strategies, called \emph{dominant strategies} in \cite{DF14}, and show that for a non-distributed system, a dominant strategy can be found (or its non-existence proved) in 2EXPTIME. That is, dominant strategies are not harder to find than the usual winning strategies. Since a dominant strategy must be winning if a winning strategy exists, this means we can find best-effort strategies in the same time complexity as usual winning strategies, without sacrificing the basic goal of correctness.

The focus of \cite{DF14} is however on the synthesis of dominant strategies for
systems with multiple processes, which is shown to be effectively decidable
(with a much lower complexity than with other approaches) for specifications
that are known to have dominant strategies. Moreover, the constructed strategies
are modular, and synthesis can even be made compositional for safety properties.
Thus, in this case we not only obtain strategies that do their best even if the
specification cannot always be fulfilled, but we can find such a strategy even
in cases where the classical distributed synthesis problem is undecidable.

Even though it is in some sense orthogonal to our question of how to
properly treat assumptions, we view the behavior of the system on lost
states as an important ingredient to building desirable systems. In a
system composed of components that are not necessarily adversarial,
this approach may help reach a common goal.
While robust synthesis attempts to satisfy guarantees as well as possible under
the worst-case environment, the best-effort strategies attempt to increase the
chances of satisfying all guarantees under a friendly environment assumption.
Both views have their merits.


\subsection{Fallback to Human}
\label{sec:hil}

\noindent
Another interesting way of dealing with ``hopeless'' situations in synthesis
has recently been presented by Li et al.~\cite{li14}.  Safety critical control
systems like autopilots in a plane or driving assistance in a car are usually
not fully autonomous but involve human operators.  If the environment behaves
such that guarantees cannot be enforced, the controller can therefore simply ask
the human operator for intervention.  This allows for semi-autonomous
controllers, even for unrealizable specifications.  There are two additional
requirements: the human operator should be notified ahead of time, and no
unnecessary intervention should be required.

The approach computes a non-deterministic counterstrategy.  In operation, the
controller constantly monitors the behavior of the environment and tracks if it
conforms to this counterstrategy.  This prevents alarms when the environment is
not fully adversarial, so that the guarantees can be enforced even though the
specification is unrealizable in the worst case. Notifying the human operator
ahead of a potential specification violation is achieved by requiring a minimum
distance (number of steps) to any failure-prone state.

The faithfulness of this approach is similar in spirit to the best-effort
strategies discussed in the previous section:  the specification cannot be
satisfied in the worst case, but this should not be an excuse for resigning. The
worst case may not occur (often) in operation, and the synthesized system should
take advantage of this.
While requiring human intervention may only be an option in specific settings,
the idea of checking the actual environment behavior against a counterstrategy
in order to assess whether the environment is behaving in an adversarial manner
is definitely interesting.


\subsection{Markov Decision Processes}
\label{sec:mdp}
Another way of refraining from worst case assumptions in synthesis is by using
Markov Decision Processes (MDPs)~\cite{Bianco95,Bollig04}. The environment is
not considered to behave adversarially but randomly with a certain probability
distribution.  This situation is also referred to as \emph{1.5 player game} (the
probabilistic environment only counts as half a player). Strategies for such
games attempt to maximize the probability to satisfy the goal.  There also exist
solutions to maximize quantitative objectives against a random
player~\cite{Chatterjee10}.

MDPs as the sole synthesis algorithm may not be satisfactory since optimality
against a random player does not necessarily imply that the strategy is winning
against an adversarial player~\cite{Faella08}. Nevertheless, MDPs can be
valuable to optimize the behavior in lost states, or to specialize a winning
strategy that allows for multiple options in several situations. 

\section{Cooperate!}
\label{sec:cooperate}

Realistic applications of synthesis methods will in general not synthesize a 
complete system from scratch, but will separate the system into components 
that can be implemented (either by hand or by synthesis) modularly. 
To make such an approach tractable while still giving global correctness guarantees, synthesis of every component must take into account the expected behavior of the rest of the system, again expressed as some kind of environment 
assumption. 

Thus far, we have discussed synthesis approaches that are designed to prefer cases where $\Gua$ is satisfied over cases where $\Ass$ is violated (Sect.~\ref{sec:lazy}), and that try to optimize the result even if the goal cannot be reached in all cases (Sect.~\ref{sec:giveup}). In some sense, the latter can be seen as an implicit collaboration with the environment, i.e., \emph{hoping} that it is not its main goal to hurt us. 

In this section, we consider synthesis algorithms for systems that 
\emph{explicitly} cooperate. In this case, 
the environment can really be considered as a second system player, and the 
payoff matrix is notably different, see Figure~\ref{fig:matrixcoop}. In 
particular, we do not want ``our'' system component to force assumption 
violations in other system components, as this would lead to incorrect 
behavior of the overall system. Instead, we want synthesis to 
be based on a ``good neighbor assumption'', i.e., the environment will only violate the 
assumptions if necessary, and we should not force it to do so, but try to 
make the overall system work even if the assumptions are not always satisfied.

\begin{figure}
\centering
\begin{tikzpicture}
\newcommand{\distance}{0.4}
\node[anchor=west] (gv) at (0,0) {$\Gua$ violated};
\node[anchor=north west] (gs) at ($(gv.south west)+(0,-\distance)$) {$\Gua$ satisfied};
\node[anchor=south west] (av) at ($(gv.north east)+(\distance,\distance)$) {$\Ass$ violated};
\node[anchor=west] (as) at ($(av.east)+(\distance,0)$) {$\Ass$ satisfied};

\path ($(gs.north west)+(-0.5*\distance,0.5*\distance)$) node (line3start) {} -| node (line3mid) {} ($(as.north west)+(-0.5*\distance,0.5*\distance)$) node (line3end) {};

\path ($(gv.north west)+(-0.5*\distance,0.5*\distance)$) node (line1start) {} -- ($(as.south east)+(0.5*\distance,-0.5*\distance)$) node (line1end) {};
\path ($(av.north west)+(-0.5*\distance,0.5*\distance)$)  node (line2start) {}  -- ($(gs.south east)+(0.5*\distance,-0.5*\distance)$) node (line2end) {};

\path (line1start) -| node (Q1OL) {} (line2start);
\path (line2start) |- node (Q1UL) {} (line3mid);
\path (line2end) -| node (Q3UR) {} (line3mid);
\path (line3start) -| node (Q2LR) {} (line1end);
\path (line1end) |- node (Q4LR) {} (line2end);

\path[line width=0pt,fill=black!20!white] (line2end) rectangle (line3mid);

\path[line width=0pt,fill=black] (Q1OL) rectangle (line3mid);

\path[line width=0pt,fill=black!50!white] (line3mid) rectangle (line1end);



\draw[thick] ($(gv.north west)+(-0.5*\distance,0.5*\distance)$) -- ($(as.south east)+(0.5*\distance,-0.5*\distance)$);
\draw[thick] ($(av.north west)+(-0.5*\distance,0.5*\distance)$) -- ($(gs.south east)+(0.5*\distance,-0.5*\distance)$);

\path (gv) -| node {\color{white}\textbf{0}} (av);
\path (gv) -| node {\color{white}\textbf{0.25}} (as);
\path (gs) -| node {\textbf{0.75}} (av);
\path (gs) -| node {\textbf{1}} (as);
\end{tikzpicture}\caption{Cooperative desirability matrix.}
\label{fig:matrixcoop}
\end{figure}

The basic idea is that environment and system can cooperate to 
some extent, in order to satisfy both $\Ass$ and $\Gua$. If we allow \emph{
full 
cooperation}, then the synthesis problem becomes the problem of synthesizing 
an implementation for both the environment and the system, and requiring them 
to jointly satisfy $\Ass \land \Gua$. This problem has been considered for different models of communication~\cite{ClarkeE81,PnueliR90,FinkbeinerS05}. Such solutions are however unsatisfactory for two reasons: 
\begin{enumerate}[(i)]
\item The approaches synthesize one particular implementation of the 
environment. This will only be a correct implementation in the overall system 
if $\Ass$ contains \emph{all} of the required properties of the rest of the 
system, not allowing us to abstract from parts of the environment. 
\item The synthesized implementation of the system is guaranteed to satisfy $
\Gua$ only for \emph{exactly this environment}. Thus, the approach does also 
not allow additional refinement or modification of the environment behavior.
\end{enumerate}
Together, these two properties imply that we cannot use such an approach to 
modularize synthesis, as we need to synthesize both components in full detail 
at the same time.

In the following, we consider \emph{assume-guarantee synthesis} 
(Sect.~\ref{sec:AGS}) and \emph{synthesis under rationality assumptions} 
(Sect.~\ref{sec:rational}), two 
approaches that are between a completely adversarial and a completely 
cooperative environment behavior. Both are based on the notion of 
non-zero-sum games, i.e., games in which players do not have mutually exclusive 
objectives, but can reach (part of) their respective objectives by cooperation.

\subsection{Assume-Guarantee Synthesis}
\label{sec:AGS}
 Intuitively, the 
\emph{assume-guarantee synthesis} approach by Chatterjee and 
Henzinger~\cite{Chatter07} 
wants to synthesize implementations for two parallel processes $P_1,P_2$ 
(which could be the system and the environment) such that solutions are robust 
with respect to changes in the other process, as long as it does not violate 
its own specification. More formally, we want to find implementations of $P_1
, P_2$ that satisfy $\spec_1 \land \spec_2$ together, and furthermore the 
solutions should be such that each process $P_i$ satisfies $\spec_j \rightarrow
 \spec_i$ for \emph{any} implementation of the other process $P_j$. That is, 
given a pair of solutions for $P_1, P_2$, we can replace one of them with a 
different implementation. As long as it satisfies its own specification $\spec
_j$ (together with the fixed implementation for the other process), we know 
that the overall specification $\spec_1 \land \spec_2$ will still hold. 

This means that players have to cooperate to find a common solution, but 
cooperation is also limited, in that the players cannot 
decide on one particular strategy to satisfy the joint specification.
Thus, assume-guarantee synthesis is an option \emph{between} purely 
adversarial and purely cooperational synthesis: if we obtain process 
implementations $P_1$ and $P_2$ that satisfy $P_i \models \phi_j \rightarrow 
\phi_i$ for adversarial synthesis, then the parallel composition $P_1 
\parallel P_2$ of these two implementations will also satisfy the conjunction 
$\phi_1 \land \phi_2$. Since each of them satisfy their spec in an arbitrary 
environment, they in particular satisfy the assume-guarantee specification. 
Moreover, every solution for assume-guarantee synthesis obviously is also a 
solution for cooperative synthesis.

\begin{example}
Consider two processes $P_1,P_2$, each with one output $o_i$ that can be read by
the other process, and specifications
\[
\phi_1 = \left\{
\begin{array}{rl}
& \always \eventually o_1\\ \land & \always \left( o_1 \rightarrow \nextt \neg o_1 \right)
\end{array}\right\},
\quad 
\phi_2 = \always \left( (\nextt o_2) \leftrightarrow o_1 \right).
\]

There are several implementations for $P_1$ that satisfy $\phi_1$ (and do 
not depend on the implementation of $P_2$), and several implementations for
$P_2$ that satisfy $\phi_2$, most of them depending on the implementation of $
P_1$. For example, $P_1$ might raise $o_1$ in the initial state, and then 
every third tick. For this implementation, a suitable implementation for $P_2$
 can raise $o_2$ in the first tick after the initial state, and then every third 
tick from there.

While this implementation for $P_2$ is correct for the particular 
implementation of $P_1$, it is not correct for all implementations of $P_1$ 
that satisfy $\phi_1$. For example, $P_1$ could raise $o_1$ every second 
tick, and the given $P_2$ would not satisfy $\phi_2$ anymore. However, there 
is an implementation that satisfies $\phi_2$ for all implementations of $P_1$ 
that satisfy $\phi_1$: $P_2$ can simply read $o_1$ and go to a state where it
raises $o_2$ iff $o_1$ it currently active. Only such a solution for $P_2$ 
solves the assume-guarantee synthesis problem (any solution for $P_1$ that 
satisfies $\phi_1$ is fine, since it does not depend on $P_2$).

Furthermore, consider the extended specification 
\[
\phi_1 = \left\{
\begin{array}{rl}
& \always \eventually o_1\\ \land & \always \left( o_1 \rightarrow \nextt \neg o_1 \right)
\end{array}\right\},
\quad 
\phi_2 = \left\{
\begin{array}{rl}
& \always \left( (\nextt o_2) \leftrightarrow o_1 \right)\\ 
\land & \always \left( \neg o_2 \rightarrow \nextt o_2 \right)
\end{array}\right\}.
\]

Now, while there are implementations for both processes with $(P_1 \parallel P_2) \models \phi_1 \land \phi_2$, there is no solution of the assume-guarantee synthesis problem: a solution for $P_2$ \emph{must} raise $o_2$ at least every second tick now, and will only work with such implementations of $P_1$, but not with those that raise $o_1$ less frequently (even if they still satisfy $\phi_1$). \exampleend
\end{example}

\subsection{Synthesis under Rationality Assumptions}
\label{sec:rational}
A number of different approaches to the synthesis of multi-component systems relies on the notion of \emph{rationality}. Informally, this means that every component has a goal that it wants to achieve, or a payoff it wants to maximize, and it will always use a strategy that maximizes its own payoff. Both the rationality of players and the goals of all components are assumed to be common knowledge. In particular, a player will only use a strategy that hurts other components if this will not lead to a smaller payoff for itself. As can be expected, this leads to implementations that do not behave purely adversarial, but cooperate to some degree in order to satisfy their own specification.

We survey three different approaches based on rationality: \emph{rational synthesis} by Fisman, Kupferman and Lustig~\cite{Fisman10}, methods based on \emph{iterated admissibility} by Berwanger~\cite{Berwanger07} and by Bernguier, Raskin and Sassolas~\cite{BrenguierRS14}, and an extension of the notion of secure equilibria to the multi-player case, called \emph{doomsday equilibria}~\cite{Chatterjee0FR14}.

\paragraph{Rational Synthesis.} The
\emph{rational synthesis} approach centers synthesis around a special \emph{system process}, and produces not only an implementation for
the system, but also strategies for all components in the environment, such that the
specification of the system is satisfied, and the strategies of the components are
optimal in some sense. To guarantee correctness, the approach assumes that these
strategies can be communicated to the other components, and that the components will not use a different strategy than the one proposed, as long as it is optimal. 

The 
definition of what is considered to be an optimal strategy leaves some freedom to the approach. The 
authors explore Nash equilibria (cp. Ummels~\cite{Ummels06}), dominant strategies (cp. Faella~\cite{Faella08}, Damm and Finkbeiner~\cite{DF14}), and subgame-prefect 
Nash equilibria (also~\cite{Ummels06}). Intuitively,
\begin{itemize}
\item if the set of proposed strategies is a Nash equilibrium profile, then 
no process can achieve a better result if it changes its strategy (while all 
others keep their strategies);
\item if the set of strategies is a dominant strategy profile, then no 
process can achieve a better result if any number of processes (including 
itself) change their strategy;
\item if the set of strategies is a subgame-perfect equilibrium profile, then 
no process can achieve a better result for any arbitrary history of the game\footnote{even those that do not correspond to the given strategy profile} by changing its strategy (while all others keep their strategies).
\end{itemize}

Compared to assume-guarantee synthesis, this approach does not guarantee that 
the synthesized 
implementation will also work when other processes change their behavior, 
even if the different behavior still satisfies the specification. Instead,
it is based on the assumption that other processes have no incentive to 
change their behavior, which is somewhat unsatisfactory for a modular
synthesis approach. 

\begin{example}
Consider again the example from above, 
\[
\phi_1 = \left\{
\begin{array}{rl}
& \always \eventually o_1\\ \land & \always \left( o_1 \rightarrow \nextt \neg o_1 \right)
\end{array}\right\},
\quad 
\phi_2 = \left\{
\begin{array}{rl}
& \always \left( (\nextt o_2) \leftrightarrow o_1 \right)\\ 
\land & \always \left( \neg o_2 \rightarrow \nextt o_2 \right)
\end{array}\right\}.
\]

A solution that satisfies $(P_1 \parallel P_2) \models \phi_1 \land \phi_2$ is also a rational synthesis solution, for any of the notions of optimality above. However, for a Nash equilibrium, a pair of implementations for $P_1, P_2$ is also a solution if $(P_1 \parallel P_2) \models \spec_1$ and $(P_1 \parallel P_2) \models \neg \spec_2$, as long as there does not exist an implementation $P_2'$ for which $(P_1 \parallel P_2') \models \spec_2$. \exampleend
\end{example}

Rational synthesis with Nash 
equilibrium has strictly weaker conditions on implementations than 
assume-guarantee synthesis. That is, any solution 
of assume-guarantee synthesis will also be a solution for this case of 
rational synthesis, but this is not always the case in the other direction. 
Also, dominant or subgame-perfect equilibria strategy profiles will always be 
Nash equilibrium profiles, but the set of solutions seems to be incomparable 
with assume-guarantee synthesis.

A combination of assume-guarantee reasoning with rational synthesis seems 
possible: instead of requiring that the system implementation works exactly 
in the given equilibrium, it should work for any behavior of the other
processes that does not reduce their payoff, or respectively any behavior 
where they still satisfy their own specification.

\paragraph{Iterated Admissibility.}
The basic idea of iterated admissibility approaches~\cite{Berwanger07,BrenguierRS14} is similar to rational synthesis: every component has its own goal in a (non-zero-sum) game, and is assumed to be rational in that it avoids strategies that are dominated by other strategies (taking into account all possible strategies of the other players). This avoidance of dominated strategies removes some of the possible behaviors for all players. Both the rationality assumption and the full state of the game being played are assumed to be common knowledge, so every player knows which strategies the other players will eliminate. Under the new sets of possible behaviors, there may be new strategies that are dominated by others, so the process of removing dominated strategies can be iterated and repeated up to a fixpoint.

The basic notions of this class of infinite multi-player games have been defined by Berwanger~\cite{Berwanger07}. Brenguier, Raskin and Sassolas~\cite{BrenguierRS14} have recently investigated the complexity of iterated admissibility for different classes of objectives, and showed that in general it is similar to the complexity of Nash equilibria.

Compared to rational synthesis, where the system process can compute strategies for all other components and they will accept them if they are optimal, in this case there is no distinguished process. Instead, all processes compute a set of optimal (or admissible) strategies, with full information allowing all components to come to the same conclusions.

\paragraph{Doomsday Equilibria.} 
The notion of \emph{doomsday equilibria} by Chatterjee et
al.~\cite{Chatterjee0FR14} uses the rationality assumption like the two
approaches mentioned before, but takes the punishment for deviating from a
winning strategy to the extreme: a doomsday equilibrium is a strategy such that
all players satisfy their objective, and if any coalition of players deviates
from their strategy and violates the objective of at least one of the other
players, then the game is doomed, i.e., the losing player(s) have a strategy
such that none of the other players can satisfy their objective.

A distinguishing feature of doomsday equilibria is that their existence is decidable even in partial information settings, in contrast to the other existing notions of equilibria. In the case of two players, doomsday equilibria coincide with the well-known notion of secure equilibria.




\begin{table}\label{tab:overview}
\centering
\newcommand{\Yes}{\ding{52}}
\newcommand{\yes}{({\scriptsize\ding{51}})}
\renewcommand{\arraystretch}{1.5}
\renewcommand{\tabcolsep}{5mm}
\caption{Comparison of existing approaches.}
\label{tab:compare}
\begin{tabular}{lc|cccc}
\toprule
 &\rotatebox{90}{Section}
 &\rotatebox{90}{Be Correct!}
 &\rotatebox{90}{Don't be Lazy!}
 &\rotatebox{90}{Never Give Up!}
 &\rotatebox{90}{Cooperate!}\\
\midrule
$\Ass \rightarrow \Gua$
&\ref{sec:standard}                 &\Yes&    &    &   \\
Strict Realizability
&\ref{sec:gr1}                      &\Yes&    &    &\yes\\
Error-Resilience / Recovery Transitions
&\ref{sec:error_res}                &\Yes&\yes&    &    \\
Robustness
&\ref{subsec:robust}                &\Yes&\Yes&    &    \\
Quantitative Synthesis
&\ref{sec:quant}                    &\yes&\Yes&    &    \\
Synthesizing Environment Assumptions
&\ref{sec:environment-assumptions}  &\Yes&    &\Yes&    \\
Best Effort Strategies
&\ref{sec:best_effort}              &\Yes&    &\Yes&    \\
Fallback to Human Control
&\ref{sec:hil}                      &\yes&    &\Yes&    \\
Markov Decision Processes
&\ref{sec:mdp}                      &    &    &\Yes&    \\
Assume-Guarantee Synthesis
&\ref{sec:AGS}                      &\Yes&    &    &\Yes\\
Synthesis under Rationality Assumptions
&\ref{sec:rational}                 &    &\yes&\yes&\Yes\\
\bottomrule
\end{tabular}
\end{table}

\section{Conclusions}
\label{sec:concl}

In this paper, we discussed the role of environment assumptions in synthesis of
reactive systems, and how existing approaches handle such assumptions. Besides
correctness, we proposed three more properties that a good system should realize:
systems should satisfy guarantees as well as possible even if environment
assumptions are violated (\emph{Don't be Lazy!}), they should aim for satisfying
the guarantees even if this is not possible in the worst case (\emph{Never Give
Up!}), and systems should rather help the environment satisfy the assumptions
instead of trying to enforce their violation (\emph{Cooperate!}).  These
properties are especially important in modular synthesis, where assumptions are
used to abstract other parts of the system rather than expressing ``don't
care''-situations. As summarized in Table~\ref{tab:compare}, we conclude that
none of the existing approaches satisfies all these requirements.  Although
important steps towards synthesis of high quality systems have been made, we
believe that even better results can be achieved by combining and extending
ideas from the different branches.  \emph{The} perfect solution may not exist,
since it may strongly depend on the application.  Even if it does exist, it may
be prohibitively expensive to achieve. In any case, more research is needed to
explore both the most important objectives and the best possible solutions.


\bibliographystyle{eptcs}
\bibliography{synthesis}
\end{document}